\documentstyle[aps,prl,multicol,graphicx]{revtex}
\begin{document}

\title{Microwave properties of $(Pr_xY_{1-x})Ba_2Cu_3O_{7-\delta }$ :
  Influence of magnetic scattering} 

\author{H.~Srikanth and S.~Sridhar}

\address{Physics Department, Northeastern University, Boston, MA 02115} 

\author{D.~A. Gajewski and M.~B. Maple} 

\address{Department of Physics and Institute for Pure and Applied Physical Sciences, University of California at San Diego, La
  Jolla, CA 92093} 

\date{\today}
\maketitle

\begin{abstract}
We report measurements of the surface impedance $Z_s=R_s+iX_s$ of $(Pr_xY_{1-x})Ba_2Cu_3O_{7-\delta}$, $(x=0,0.15,0.23,0.3,0.4,0.5)$. Increasing $Pr$ concentration leads to some striking results not observed in samples
doped by non-magnetic constituents. The three principal features of the $R_s(T)$
data - multiple structure in the transition, a high residual resistance and,
at high $Pr$ concentrations, an upturn of the low $T$ data, are all
characteristic of the influence of magnetic scattering on superconductivity,
and appear to be common to materials where magnetism and superconductivity
coexist. The low $T$ behavior of $\lambda (T)$ appears to change from $T$ to 
$T^4$ at large $Pr$ doping, unlike that reported for $Ni$ and $Zn$ substitutions, and is further
evidence of the influence of magnetic pairbreaking of the $Pr$.
\end{abstract}

Keywords: microwave absorption, penetration depth, pair breaking

\begin{multicols}{2}
In the 123 class of cuprates where Y is replaced by lanthanide elements 
such as La,Ce,Pr,Nd,Gd,Dy, etc., the compound $PrBa_2Cu_3O_{7-\delta }$ is 
insulating while all the other members of the family show a superconducting 
transition in the vicinity of $90K$. 
Superconductivity in $(Pr_xY_{1-x})Ba_2Cu_3O_{7-\delta }$ is suppressed 
rapidly as the $Pr$ content is increased and the system undergoes a transition
 to an insulating state at $x\sim 0.55-0.6$ \cite{HBRadousky92a,MBMaple94d}. Since superconductivity in the cuprates is mainly associated with the $Cu-O$ 
planes, the exact role of $Pr$ (which substitutes for $Y$ in the $123$ 
structure) in the $T_c$ suppression is of fundamental interest. This also 
makes the $Pr$ doped system distinct from the other transition metal doped 
$123$ compounds where dopants like $Ni,Zn,Fe$ are substituted at the $Cu$ 
sites and thus directly affect the superconductivity in the planes. 
Two mechanisms for the decrease of $T_c$ with $Pr$ concentration $x$ have been 
proposed: (1) annihilation of mobile holes in the $CuO_2$ planes by the $Pr$ 
ions (hole depletion mechanism), and (2) superconducting electron 
pair breaking (pair breaking mechanism) \cite{MBMaple94d}. Superconducting 
electron pair breaking could be produced by potential scattering of mobile 
holes by the $Pr$ ions if $YBa_2Cu_3O_{7-\delta}$ is a $d$-wave superconductor
 and by spin-dependent exchange scattering of the mobile holes by the $Pr$ 
ions, which carry well-defined magnetic moments, if $YBa_2Cu_3O_{7-\delta}$ 
is an s-or d-wave (spin-singlet) superconductor \cite{DLCox95a}. Both 
of these mechanisms have been incorporated into a phenomenological model which
 provides a semiquantitative description of the striking variations of $T_c$ 
with $x$ and $y$ in the $(Ca_xPr_yY_{1-x-y})Ba_2Cu_3O_{7-\delta}$ system for 
$0 \le x, y \le 0.2$ (here, mobile holes are generated by $Ca$ and annihilated
 by $Pr$ \cite{MBMaple94d,JJNeumeier89} and the pressure dependence of $T_c$ 
in the $(Pr_xY_{1-x})Ba_2Cu_3O_{7-\delta}$ system for $0 \le x \le 0.5$ \cite{JJNeumeier88,MBMaple92}. The hole-depletion and pair breaking mechanisms are 
assumed to originate in the hybridization of the localized $Pr 4f$ states and 
the $CuO_2$ valence band states. The existence of $Pr 4f - CuO_2$ valence band
 hybridization was first proposed on experimental grounds to account for the 
anomalous pressure dependence of $T_c$ of  $(Pr_xY_{1-x})Ba_2Cu_3O_{7-\delta}$ 
system \cite{JJNeumeier88,MBMaple92}, in analogy with the anomalous behavior 
of $T_c$ under pressure in superconductors containing $Ce$ impurities such as 
$La_{1-x}Ce_x$ \cite{MBMaple76}. To the extent that pair breaking is 
responsible for the depression of $T_c$ in the $(Pr_xY_{1-x})Ba_2Cu_3O_{7-\delta}$ system, it would be necessary to invoke $Pr 4f-CuO_2$ valence band hybridization in 
order to generate a sufficiently strong coupling of the $Pr$ ions to the 
mobile holes in the $CuO_2$ planes. Except for $Ce$, the other lanthanide 
(Ln) ions with partially-filled $4f$ electron shells do not depress the $T_c$ 
of $YBa_2Cu_3O_{7-\delta}$ by a measurable amount; evidently, the Ln 
$4f-CuO_2$ valence band hybridization and, in turn, the exchange coupling of
 the Ln ions to the holes in the $CuO_2$ planes is small. It is interesting to note that such a hybridization-induced exchange interaction is antiferromagnetic and should produce a Kondo effect, resulting in a depression of $T_c$ with 
impurity concentration (here, $Pr$) that deviates from the prediction of the 
theory of Abrikosov and Gor'kov (AG) in a manner that depends on the ratio of 
the Kondo temperature $T_K$ to the $T_c$ of the host superconductor (in this 
case, YBCO) \cite{MBMaple76}. Measurements of the low temperature specific 
heat in the range $0 < x \le 0.5$ reveal a broad anomaly in the specific heat 
\cite{AKebede89,SGhamaty91}. This anomaly can be described by the sum of a 
term of the form $C_e(T) = \gamma T$ with an enormous ``heavy-fermion-like'' value of $\gamma$ of $\sim 240 mJ/mol Pr-K^2$ and a contribution $C_M(T)$ that has been described by a spin 1/2 Kondo anomaly with a value of $T_K$ that increases with $x$ \cite{SGhamaty91} or an anomaly associated with antiferromagnetic 
ordering of the $Pr$ ions \cite{NEPhillips91}. It has been shown that the 
detailed $T_c$ vs $x$ curve of the $(Pr_xY_{1-x})Ba_2Cu_3O_{7-\delta}$ system 
does not conform to the AG theory, but can be described by the aforementioned 
pheomenological model based on both hole depletion and pair breaking in the 
range $0 \le x \le 0.2$ \cite{JJNeumeier92}. Many experiments have indicated 
$Pr$ to be predominantly in a 3+ valence state, although a mixed valence state
 cannot be ruled out.

There are several anomalies observed in transport and magnetization experiments which indicate that $Pr_xY_{1-x}Ba_2Cu_3O_{7-\delta }$ has a ground state with unusual electronic properties, and which has been proposed theoretically to 
account for the insulating nature of $PrBa_2Cu_3O_{7-\delta}$ \cite{RFehrenbacher93a}. Neutron scattering experiments indicate long range $AFM$ ordering of 
the $Pr$ moments in the insulating compound $PrBa_2Cu_3O_{7-\delta }$ with an 
unusually high Ne\'{e}l temperature $T_N$ of $17K$ \cite{HBRadousky92a}. 
Overall the $Pr_xY_{1-x}Ba_2Cu_3O_{7-\delta}$ system represents the ideal 
candidate to explore the effects of pair-breaking and magnetic scattering in 
the superconducting state.

Microwave experiments have been shown to be unique probes of the superconducting state. Measurements of the $T$ dependence of the microwave surface impedance
 $Z_s$ and penetration depth $\lambda $ can reveal important information 
regarding the gap parameter and the quasiparticle density of states. The 
linear behavior of $\lambda $ at low $T$ observed in YBCO \cite{WNHardy93a} 
and BSCCO\cite{TJacobs95d} is consistent with nodes in the in-plane gap. It is
 of interest to examine the role of impurities in this context, since the 
influence of impurities are expected to be strikingly different for a d-wave 
superconductor compared to the conventional s-wave type as discussed above.

In this paper, we report on complex surface impedance ($Z_s=R_s+iX_s$) measurements on a series of $Pr_xY_{1-x}Ba_2Cu_3O_{7-\delta }$ single crystals with 
$x$ ranging from $0$ to $0.5$. Our experiments reveal novel structure in the superconducting transition region not seen in $dc$ or low frequency probes of 
resistivity and magnetization. In the following sections we present the surface resistance $R_s(T)$ and the low temperature penetration depth $\lambda (T)$ 
data and discuss our results in terms of enhanced magnetic scattering.

\section{Experiment}

The single crystals were grown using a flux-growth method described in an earlier
 publication \cite{LMPaulius94}. The $Pr$ concentration $x$ of the single crystals was inferred from their measured $T_c's$ and the $T_c$ vs $x$ curve of 
polycrystalline  $Pr_xY_{1-x}Ba_2Cu_3O_{7-\delta }$ samples reported in ref. 
\cite{JJNeumeier92}.  Typical crystals used in our experiments were of size
 $0.7$ x $0.7$ x $0.05$ $mm^3$. Surface impedance measurements were done in a 
superconducting $Nb$ cavity resonator operating at a frequency of $10$ $GHz$. 
The samples were mounted on a sapphire rod and a ''hot finger'' technique was
 employed to monitor the temperature dependence of the complex surface 
impedance from $4K$ to $300K$\cite{SSridhar88}. This cavity perturbation 
method has been extensively validated in precision measurements of $R_s$ and 
$X_s$ in single crystals of cuprate\cite{TJacobs95d,HSrikanth96,SSridhar97} and borocarbide superconductors\cite{TJacobs95b}. The $Nb$ cavity is maintained either at $4.2K$ or below $2K$ and the 
typical background $Q_b$ of the cavity can be as high as $10^8$. The surface 
resistance $R_s(T)$ is measured from the temperature dependent $Q$ using 
$R_s(T)=\Gamma [Q^{-1}(T)-Q_b^{-1}(T)]$ and the penetration depth using 
$\Delta \lambda (T)=\zeta [f(T)-f_b(T)]$ where the geometric factors are 
determined by the cavity mode, crystal size and location within the cavity. In
 the present setup, all measurements were done in the $TE_{011}$ mode with the
 sample at the midpoint of the cavity axis where the microwave magnetic fields
 have a maximum and the microwave electric fields are zero. In all cases, the 
samples were oriented with $H_{rf}\Vert c$ and currents only flow
in the $ab-plane$.

\section{Results and Discussion}

\begin{figure}[htbp]
\narrowtext
\begin{center}
  \includegraphics*[width=0.45\textwidth]{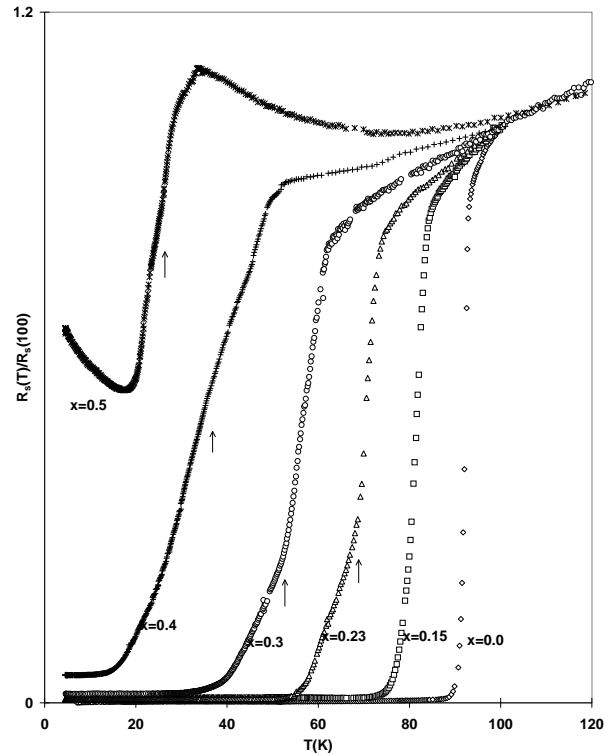}
  \caption{Surface Resistance of $Pr_xY_{1-x}Ba_2Cu_3O_{7-\delta}$. 
    The superconducting $T_c's$ for x = 0.0, 0.15, 0.23, 0.3, 0.4, 0.5
    are 92K, 82K, 71K, 58K, 46K, and 22K, respectively.}
  \label{Fig1}
\end{center}
\end{figure}

In Fig.\ref{Fig1}, the surface resistance $R_s(T)$ for $Pr_xY_{1-x}Ba_2Cu_3O_{7-\delta }$ with $x=0.0,0.15,0.23,0.3,0.4,0.5$ are shown. The data are normalized at $R_s(100K)$ to highlight the systematic variation of the surface resistance with increasing $Pr$ concentration. The normal state $R_s(100K)$ values 
range from $0.2$ to $0.8$ $\Omega $ as $x\rightarrow 0$ to $0.5$. The overall 
temperature dependence of the surface impedance in the normal state is consistent with the expected skin depth limited response given by $R_n=\sqrt{\omega \mu _0\rho _n/2}$ where $\rho _n$ is the normal state $dc$ electrical resistivity. This is also well represented in the upturn in $R_n$ clearly seen for the 
$x=0.5$ sample before the superconducting transition occurs. At $x=0.5$, the 
$Pr$-doped system is on the verge of a metal-insulator transition and $dc$ electrical resistivity measurements indicate that this transition occurs for 
$x\sim 0.55-0.6$. This transition from a linear normal state resistivity in 
the metallic case to a Mott Variable Range Hopping (VRH) type behavior in the insulating regime is mirrored in the surface resistance
data of Fig.\ref{Fig1}. In particular the change in sign of $dR_s/dT$ from 
positive to negative just above the superconducting transition is evident as 
$x$ increases from $0.4$ to $0.5$. The transition is sharpest for the undoped
 $YBCO$ sample with a width $\Delta T_c$ which increases rapidly with $Pr$ 
substitution. Random substitution of $Y$ atoms by $Pr$ which increases the 
disorder in the system combined with the magnetic scattering due to free $Pr$ 
ions is the likely cause for the broad transition.

An interesting feature of the data of Fig.\ref{Fig1} is the remarkable fine
structure seen in the transition region. This is particularly accentuated in
the $R_s$ data for $x=0.23$ and $x=0.3$ (marked by arrows). A distinct change
in slope at a characteristic temperature close to midpoint of the transition
width occurs and is reproducible in several $Pr$-doped crystals
investigated. It is to be noted that for the same concentrations, the $dc$
resistivity and susceptibility measurements do not show any signature of
multiple features in the transition. We propose that this two-slope
structure is a consequence of a discontinuous change in quasiparticle
scattering in the superconducting state at characteristic temperatures. This
is seen clearly in our high frequency experiments as we probe both the
normal and superfluid response in the superconducting state.

The residual surface resistance at low temperatures shows an increasing
trend with $Pr$ doping, and there is almost an order of magnitude jump
between $x=0.4$ and $0.5$. The most striking aspect of the $x=0.5$ sample is
the upturn in $R_s$ for $T\leq 18K$ and the temperature dependence almost
mimics the normal state data just above the superconducting transition. This
is an important observation which we believe for the first time
qualitatively captures the competing effects of magnetism and
superconductivity in cuprates as revealed in surface impedance measurements. It should also be pointed out that the presence of exchange interactions due to the $Pr4f$ moments could also result in enhanced electron-electron interaction effects. In this case, many body corrections to the low temperature quasiparticle conductivity have to be taken into account and this can lead to the upturn in the $R_s(T)$ seen for higher $Pr$ doping.

It is to be pointed out that all the anomalous characteristics seen here for
the $Pr_xY_{1-x}Ba_2Cu_3O_{7-\delta }$ system have been observed by us in
our microwave experiments of borocarbide class of magnetic superconductors 
\cite{TJacobs95b}. In general, a high residual surface resistance, multiple
structure in the transition and the low temperature upturn in $R_s$ all seem
to be standard electrodynamic characteristics of systems exhibiting co-existence of
magnetism and superconductivity. Systematic high frequency experiments on
known magnetic superconductors would help further understand the implication
of these features.

\begin{figure}[htbp]
\narrowtext
\begin{center}
    \includegraphics*[width=0.45\textwidth]{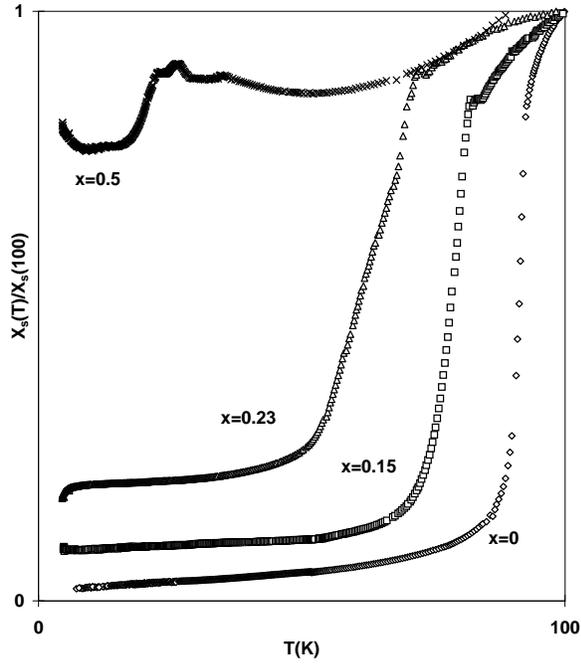}
  \caption{Surface Reactance $X_s$ of $Pr_xY_{1-x}Ba_2Cu_3O_{7-\delta}$.}
  \label{Fig2}
\end{center}
\end{figure}

\begin{figure}[htbp]
\narrowtext
 \begin{center}
    \includegraphics*[width=0.45\textwidth]{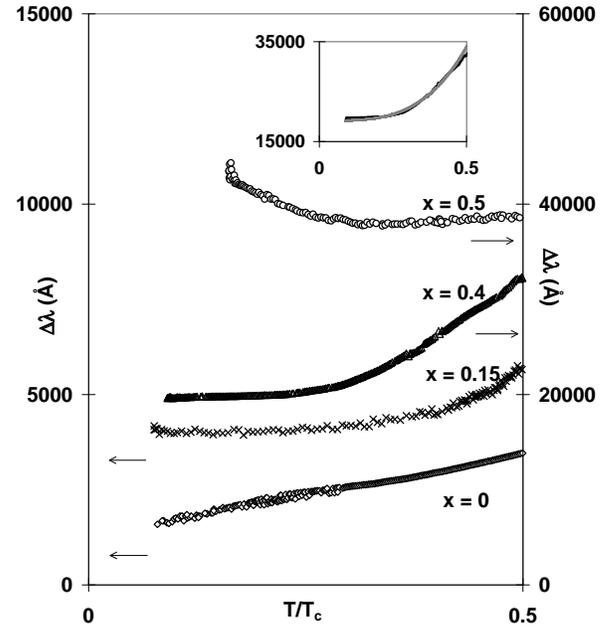}
  \caption{Low temperature penetration depth {$\Delta\lambda(T)$ $\it vs$ reduced temperature $t = T/T_c$}.
    For x = 0.15 and 0.4 the data follows a $t^4$ dependence. Inset shows the data for $x = 0.4$ (solid line) along with a $t^4$ curve (dashed line) for comparison. An
    approximate $ \lambda_0$ is added to each data.}
  \label{Fig3}
\end{center}
\end{figure}

The normalized surface reactance $X_s(T)$ is shown in Fig.\ref{Fig2}. The
overall trend is similar to the corresponding $R_s(T)$ data with some
differences particularly for the $x=0.5$ sample. Two peaks are present in
the vicinity of the superconducting transition and a third feature at least $%
15K$ higher in temperature. It is not clear whether these features are
related to coherence effects. The surface reactance also shows an upturn at
low temperature but the onset of this is about $10K$ lower than the
corresponding onset in $R_s$. The change in reactance is directly related to
the change in penetration depth via the relation $X_s=\mu _0\omega \lambda $%
. In Fig.\ref{Fig3}, the low temperature penetration depth extracted from
the surface reactance is plotted as a function of reduced temperature ($%
T/T_c $). Measurement of the absolute value of the penetration depth is not
possible with the cavity perturbation technique. However, an estimate can be
obtained by setting $R_s=X_s$ in the normal state assuming local
electrodynamics. An estimated London penetration depth $\lambda _0$ is added
to each set of data in Fig.\ref{Fig3}.

For the undoped $YBCO$ sample ($x=0$), the penetration depth is linear in
temperature. This linear dependence seen both in $YBCO$ and $BSCCO$ class of
cuprates has been considered as evidence for an order parameter symmetry
having nodes in the gap\cite{WNHardy93a,TJacobs95d}. With $Pr$ doping, the
temperature dependence of $\lambda $ changes to a power law behavior. For
the $x=0.15$ and $0.4$, the best fit to the data is obtained with a $T^4$
term. It is important to note that the data $do$ $not$ show a $T^2$
behavior reported in thin films of $YBCO$\cite{ZMa93a} and $Zn$-doped $YBCO$
single crystals\cite{DABonn94a}. The change over from $T$ to $T^2$
dependence in $Zn$ and $Ni$ doped $YBCO$ has been interpreted as due to a
crossover from a pure $d-wave$ state to a gapless superconducting state.
There are also a number of differences in $R_s$ between the $(Zn,Ni)$ and $Pr $ doped $YBCO$. In the case of $Zn$ or $Ni$ doped samples, although the
normal state $R_s$ increases with doping, there is negligible change in the
low temperature residual surface resistance. On the contrary, the normal
state and residual $R_s$ increase with $Pr$ concentration in $Pr_xY_{1-x}Ba_2Cu_3O_{7-\delta }$. No multiple structure is seen in the
transition for $Zn$ or $Ni$ doped samples. Thus there is every indication
that enhanced scattering presumably of magnetic origin plays a major role in
the surface impedance characteristics of $Pr$ doped $YBCO$.

In a recent paper, it has been theoretically proposed that in layered
superconductors it is possible to have a novel phase transition governed
entirely by the scattering rate $1/\tau $ provided the order parameter
reverses its sign on the Fermi surface but its angular average is finite\cite
{SVPokrovsky95a}. According to this, the excitation energy spectrum which is
gapless at a low level of scattering can develop a gap as the scattering
rate exceeds some critical value; i.e., $1/\tau >1/\tau ^{*}$. Our surface
impedance data on the $Pr$ doped crystals seem to be remarkably consistent
with this scenario. With increasing $Pr$ concentration, not only are we
introducing pair-breaking magnetic ions but also increasing the scattering
rate. While the $T_c$ suppression can be thought of as due to hole depletion and magnetic
pair-breaking, the enhanced scattering rate may result in the kind of
transition proposed by this theory. Both the two-slope feature in the
transition region and change in the temperature dependence of the
penetration depth from linear to power-law shown in Fig.\ref{Fig1} and Fig.%
\ref{Fig3} can be reconciled as a manifestation of a transition from a
gapless state to a superconducting state with a finite gap.

Finally, we would like to remark about the case of mixed state order parameter symmetry like $s+d$ in the 123 system. In this case, where the order parameter has line nodes on the Fermi surface, elastic scattering suppresses the $T_c$ vigorously and has a dominant influence over spin-flip scattering. The transition for a linear $T$ to $T^4$ dependence in $\lambda$ can be interpreted as the suppression of the d-part of the order parameter, whereas insensitive to the impurities, s-wave survives. It is to be noted that our recent microwave experiments on high quality $YBa_2Cu_3O_{7-\delta}$ single crystals grown in $BaZrO_3$ crucibles indicate evidence for a multi-component order parameter symmetry \cite{HSrikanth96,SSridhar97}.  

In conclusion, we have presented the microwave surface impedance for $(Pr_xY_{1-x})Ba_2Cu_3O_{7-\delta }$ single crystals with $x$ ranging from $0$ to 
$0.5$. Both the surface resistance and penetration depth data indicate
strong scattering effects. Novel structure in the superconducting transition
region and the temperature dependence of the penetration depth may be
consistent with a phase transition from a gapless to finite-gap state
governed by the scattering rate.

\section{Acknowledgements}
Work at Northeastern was supported by NSF-DMR-9623720. The research at UCSD was supported by the U.S. Department of Energy under Grant No. DE-FG03-86ER-45230. The authors would like to thank Balam A. Willemsen and T. Jacobs for helpful discussions and Valery Pokrovsky for valuable comments.

\end{multicols}
\end{document}